\documentclass{emulateapj}

\slugcomment{ApJ, submitted}

\shorttitle{Maser disk eccentricity}
\shortauthors{Armitage}

\begin{document}

\title{Eccentricity of masing disks in Active Galactic Nuclei}

\author{Philip J. Armitage\altaffilmark{1,2}}
\altaffiltext{1}{JILA, Campus Box 440, University of Colorado, Boulder CO 80309; 
pja@jilau1.colorado.edu}
\altaffiltext{2}{Department of Astrophysical and Planetary Sciences, University of Colorado, Boulder CO 80309}

\begin{abstract}
Observations of Keplerian disks of masers in NCG~4258 and other Seyfert galaxies can be 
used to obtain geometric distance estimates and derive the Hubble constant. The 
ultimate precision of such measurements could be limited by uncertainties in the disk geometry. 
Using a time-dependent linear theory model, we study the evolution of a thin initially eccentric disk under 
conditions appropriate to sub-pc scales in Active Galactic Nuclei. The evolution is controlled by a combination 
of differential precession driven by the disk potential and propagating eccentricity waves that 
are damped by viscosity. A simple estimate yields a circularization timescale of $\tau_{\rm circ} \sim 
10^7 (r / 0.1 \ {\rm pc})^{5/6} \ {\rm yr}$. Numerical solutions for the eccentricity 
evolution confirm that damping commences on this timescale, but show that the subsequent 
decay rate of the eccentricity depends upon the uncertain strength of viscous damping of 
eccentricity. If eccentricity 
waves are important further decay of the eccentricity can be slow, with full circularization requiring 
up to 50~Myr for disks at radii of 0.1~pc to 0.2~pc. Observationally, this implies that 
it is plausible that enough time has elapsed for the eccentricity of masing disks to have 
been substantially damped, but that it may not be justified to assume vanishing eccentricity. 
We predict that during the damping phase the pericenter of the eccentric orbits describes a 
moderately tightly wound spiral with radius. 
\end{abstract}

\keywords{accretion, accretion disks --- masers --- galaxies: active --- galaxies: Seyfert --- 
galaxies: distances and redshifts --- galaxies: individual (NGC 4258)}

\section{Introduction}
Observations of water maser emission from a geometrically thin disk within the 
central pc of NGC~4258 \citep{miyoshi95} have permitted a geometric 
determination of the distance to the galaxy \citep{herrnstein99,humphreys08}, and an 
 estimate of the Hubble constant. The recent discovery of apparently 
similar maser emission in a number of more distant galaxies has raised the 
exciting prospect of a percent level determination of $H_0$ independent 
of both distance ladder uncertainties and other cosmological parameters. 
Such a determination would have considerable cosmological utility, perhaps 
most importantly by improving the ability of current and future datasets to constrain 
the behavior of dark energy \citep{hu05,olling07}.

The precision with which $H_0$ can be determined from measurements of masers 
in a single system is limited by uncertainties in the galactic peculiar 
velocity, observational errors, and any unmodeled complexity in the geometry of 
the source. The first two of these limits present instrumental and observational  
challenges, and considerable progress is guaranteed given sensitive, multi-epoch 
observations of maser disks at higher redshift than the one in NGC~4258. The third 
requires both observations and theory. Typically, it is assumed that the 
maser emission originates from a thin, circular disk, and this simplest 
assumption is consistent with all existing data on NGC~4258 \citep{humphreys08}. 
There is no obvious theoretical rationale, however, for assuming that the eccentricity is 
exactly zero, or for extrapolating the NGC~4258 results to other systems. Indeed, 
the observation that the masing disk in NGC~4258 is measurably warped  
suggests the possibility that different disks might exhibit a range of geometries.
Although the geometry can be constrained observationally, this requires the 
introduction of additional parameters whose fitting inevitably degrades 
the precision with which the distance can 
be measured.

The purpose of this paper is to outline, from a theoretical perspective, the 
expected evolution of the eccentricity of a geometrically thin accretion disk at 
sub-pc distances from a supermassive black hole. Using a time-dependent 
linear theory model of an eccentric disk \citep{goodchild06} I examine both 
the rate at which eccentricity decays, and the transient structure of the 
eccentric disk before it becomes circular. 

\section{Description of the disk evolution}
Consider a thin, flat disk in which the vertical scale height $h \ll r$. We 
define a complex eccentricity via,
\begin{equation} 
 E \equiv e \exp\left[{i \varpi}\right],
\end{equation}
where $e$ is the magnitude of the eccentricity and $\varpi$ the longitude of 
pericenter of the fluid streamlines. We assume that at the radii of interest 
the potential is dominated by the Keplerian potential of the black hole 
$\Phi_{\rm BH}$, to which must be added an additional contribution due 
to the disk,
\begin{equation}
 \Phi = \Phi_{\rm BH} + \Phi_{\rm disk}.
\end{equation}
The fluid in the disk has density $\rho$, pressure $p$, angular velocity 
$\Omega$ and adiabatic exponent 
$\gamma$. We assume that turbulent stresses within the disk act to damp 
eccentricity, and parameterize the efficiency of the damping via a bulk 
viscosity $\alpha_e$ written in terms of a Shakura-Sunyaev (1973) $\alpha$ parameter.
Linearization of the two-dimensional fluid equations (in the inviscid limit, 
except for the aforementioned bulk viscosity) yields an evolution equation 
for the eccentricity \citep{goodchild06},
\begin{eqnarray}
 2 r \Omega \frac{\partial E}{\partial t} &=& 
 -\frac{iE}{r} \frac{\partial}{\partial r} \left( r^2 \frac{{\rm d}\Phi_{\rm disk}}{{\rm d}r} \right)
 + \frac{iE}{\rho} \frac{\partial p}{\partial r} \nonumber \\
 &+& \frac{i}{r^2 \rho} \frac{\partial}{\partial r} 
 \left[ \left(\gamma-i \alpha_e \right) p r^3 \frac{\partial E}{\partial r} \right].
\label{eq_evolve} 
\end{eqnarray} 
The first two terms on the right hand side describe precession driven by any 
non-Keplerian part of the disk potential and by pressure gradients within 
the disk. These terms do not change the magnitude of the eccentricity. 
The third term has the form of a Schr\"odinger equation -- it 
describes waves of eccentricity that propagate radially through the disk 
and are damped by the action of viscosity.

Theoretically, it is expected that accretion disks in Active Galactic Nuclei (AGN) 
ought to be self-gravitating at the relatively large radii where masing occurs, and this 
permits some simplification of the evolution equation. We write the 
surface density as a power-law in radius,
\begin{equation}
 \Sigma = \Sigma_0 r^\beta
\end{equation}
and compute the disk potential $\Phi_{\rm disk}$ from the enclosed 
disk mass
\begin{equation}
 M_{\rm disk} = \int^r 2 \pi r \Sigma dr
\end{equation}
assuming a spherically symmetric mass distribution. We assume that over the 
relatively narrow radial range for which masing occurs the sound 
speed $c_s$ can be taken (approximately) to be constant, and note 
that $\rho \simeq \Sigma / 2h$ and that $h = c_s / \Omega$. The 
rate of precession due to pressure gradients is then $\propto 
c_s^2$ (a constant at the radii where masing occurs), while the 
rate of precession due to the non-Keplerian potential is 
$\propto \Sigma_0$, which rises with the disk mass\footnote{These 
terms have different signs, so it is possible to choose a 
surface density and sound speed profile for which they cancel, or 
for which the disk precesses as a rigid body. Indeed, \citet{statler01} 
derived a family of potentially long-lived eccentric disk models 
which owe their stability to rigid body precession. We do not consider 
special solutions of this type here, since there is no reason to 
expect that masing disks would have a structure that leads to 
vanishing differential precession.}. Their ratio,
\begin{equation}
 \frac{\vert \dot{E}_{\rm p} \vert}{\vert \dot{E}_{\rm grav} \vert} 
 \sim \left( \frac{h}{r} \right)^2 \frac{M_{\rm BH}}{M_{\rm disk}}
\end{equation}
therefore depends upon the disk mass -- in low mass disks precession 
will be dominated by the pressure term while in higher mass disks 
gravity will dominate. The latter limit is probably appropriate 
for masing disks. At the radii -- of the order of a tenth of a 
pc -- where the masing is observed it is likely that disks around 
supermassive black holes are self-gravitating \citep{kolychalov80,clarke88,
shlosman90,goodman03}. Self-gravitating disks develop spiral structure, 
whose existence in the NCG~4258 disk is hinted at by observations of 
clustering and asymmetry in the maser emission regions \citep{maoz95,maoz98}.  
For a self-gravitating disk $M_{\rm BH} / M_{\rm disk} \approx (h/r)^{-1}$,  
$\vert \dot{E}_{\rm p} \vert / \vert \dot{E}_{\rm grav} \vert \sim 
h/r$, and pressure effects are negligible. Writing $p = \rho c_s^2$ 
the evolution equation (\ref{eq_evolve}) then simplifies to,
\begin{eqnarray}
 2 r \Omega \frac{\partial E}{\partial t} &=&
 2 \pi (\beta+1) G \Sigma_0 r^\beta i E \nonumber \\
 &+& \frac{i c_s^2}{r^2 \rho} \frac{\partial}{\partial r} 
 \left[ \left(\gamma-i \alpha_e \right) \rho r^3 \frac{\partial E}{\partial r} \right].
\label{eq_simplified} 
\end{eqnarray}
We study the eccentricity evolution implied by this equation in \S3.  

\subsection{Limitations of this description}
Equation (\ref{eq_evolve}) is an extremely simple representation of the 
evolution of an eccentric fluid disk. The most obvious limitation is 
the use of a linear equation rather than the nonlinear formalism 
developed by \cite{ogilvie01}. The linear equation ought to provide an 
approximate description of the dynamics at late times, when the 
eccentricity is small, but will evidently fail if the initial 
eccentricity and / or its variation with radius is large. A second  
and even more important limitation concerns the uncertain 
nature of the eccentricity damping. We have parameterized the damping 
via a bulk viscosity -- rather than the usual shear viscosity used in 
the Shakura-Sunyaev theory of accretion disks -- because disks dominated 
by a Navier-Stokes shear viscosity are unstable to {\em growth} of 
eccentric modes \citep{kato83,ogilvie01}. This is probably not a physical 
problem, rather it reflects the fact that angular momentum transport 
mediated by the magnetorotational instability \citep{balbus98}, self-gravity, 
or other physical mechanisms cannot be well described via a Navier-Stokes  
shear viscosity. For our purposes we simply assume that the stress in the 
disk acts such as to damp eccentricity, parameterize that damping efficiency 
(arbitrarily) via a bulk viscosity, and treat $\alpha_e$ as a free parameter.
  
\section{Eccentricity evolution}

\subsection{Analytic estimates}
The terms in equations (\ref{eq_evolve}) and (\ref{eq_simplified}) describing precession 
do not alter the magnitude of the disk eccentricity. The wave-like term proportional 
to $\gamma$ {\em does} change the local eccentricity, but preserves a global invariant 
\citep{goodchild06},
\begin{equation}
 {\cal{E}}^2 \equiv \int_{r_{\rm in}}^{r_{\rm out}} 
 \frac{1}{2} r^3 \rho \Omega \vert E \vert^2 dr.
\label{eq_invariant} 
\end{equation}
Damping of ${\cal{E}}^2$ occurs due to the viscous term at a rate
\begin{equation}
 \frac{{\rm d} {\cal{E}}^2}{{\rm d} t} = - \int_{r_{\rm in}}^{r_{\rm out}} 
 \frac{1}{2} \alpha_e p r^3 
 \vert \frac{\partial E}{\partial r} \vert^2 dr.
\label{eq_damprate} 
\end{equation}  
The presence of eccentricity waves means that the radial profile of 
$E$ needs to be considered along with the temporal evolution. This requires a 
numerical solution of equation (\ref{eq_simplified}), which we defer to \S3.2. 
For an estimate, however, we can define a local damping timescale,
\begin{equation} 
 \tau_{\rm damp} = \frac{\rho \Omega \vert E \vert^2}{\alpha_e p \vert \partial E 
 / \partial r \vert^2},
\label{eq_tcirc} 
\end{equation}
which is simply the ratio of the right hand sides of equations (\ref{eq_invariant}) 
and (\ref{eq_damprate}). If $E$ varies on the same scale as the azimuthally 
averaged disk properties then to order of magnitude $\vert \partial E / \partial r \vert^2 
\simeq E^2 / r^2$ and we find that,
\begin{equation}
 \tau_{\rm damp} = \frac{1}{\alpha_e \Omega} \left( \frac{h}{r} \right)^{-2}.
\end{equation} 
This is just the usual ``viscous time'' of the disk if $\alpha_e$, the viscosity 
responsible for damping eccentricity, is comparable to the shear viscosity $\alpha_s$  
driving mass inflow\footnote{This argument leads to the rule of thumb that an 
eccentric disk ought to circularize ``on the viscous timescale''. This is potentially 
misleading, first because there is no direct physical relation between $\alpha_e$ and 
$\alpha_s$, and, second, because differential precession can lead 
to a situation in which $\vert \partial E / \partial r \vert^2 \gg E^2 / r^2$ in 
much less than a viscous time, allowing for faster eccentricity evolution.}. It is 
well known that, at radii of 0.1~pc and beyond, the viscous time in AGN disks 
is extremely long \citep{shlosman90}. For the specific case of NGC~4258, observations 
of the thickness of the masing disk imply a midplane temperature of $T \simeq 600 \ {\rm K}$  
and a sound speed $c_s \simeq 1.5 \ {\rm kms}^{-1}$ \citep{argon07}, consistent 
with the conditions needed for H$_2$O maser emission. Adopting these values, the 
viscous time is
\begin{eqnarray}
 \tau_{\rm visc} &=& 5.7 \times 10^8  
 \left( \frac{\alpha_s}{0.1} \right)^{-1} 
 \left( \frac{c_s}{1.5 \ {\rm kms}^{-1}} \right)^{-2} \nonumber \\
 &\times& \left( \frac{M_{\rm BH}}{4 \times 10^7 \ {\rm yr}} \right)^{1/2} 
 \left( \frac{r}{0.1 \ {\rm pc}} \right)^{1/2} \ {\rm yr}.
\end{eqnarray} 
Unless sub-pc disks in AGN remain unperturbed across timescales of the 
order of a Gyr, circularization {\em due to viscous evolution} will not occur.

Much faster circularization occurs as an indirect consequence of differential 
precession, which for the disks of interest is of the order of $(h/r)^{-1} \sim 10^3$ 
times faster than viscous evolution. The precession timescale $\tau_{\rm p} = 2\pi / \omega$, 
where 
\begin{equation} 
 \omega = - \frac{1}{2 r^2 \Omega} \frac{\rm d}{{\rm d}r} 
 \left( r^2 \frac{{\rm d} \Phi_{\rm disk}}{{\rm d} r} \right).
\end{equation} 
As noted above, we expect disks at radii of the order of 0.1~pc around AGN to be 
self-gravitating, in which case \citep{toomre64}
\begin{equation}
 Q \equiv \frac{c_s \Omega}{\pi G \Sigma} \sim 1
\end{equation}  
where $\Sigma$ is the disk surface density. Using this relation to fix the 
surface density in terms of the sound speed (assumed constant) and disk 
radius, we find,
\begin{equation}
 \tau_p = 8.2 \times 10^5 Q 
 \left( \frac{c_s}{1.5 \ {\rm kms}^{-1}} \right)^{-1}
 \left( \frac{r}{0.1 \ {\rm pc}} \right) \ {\rm yr},
\end{equation}
independent of the black hole mass. This relatively rapid precession 
has an important consequence. Although it does not alter $e$, the 
differential precession leads to a rapid variation of $\varpi$ with 
radius. This reduces the characteristic lengthscale over with $E$ 
varies in equation (\ref{eq_tcirc}), and allows viscosity to damp 
the eccentricity much more quickly.  

The origin of the gas in the masing disks of Seyfert galaxies is 
not known. One possibility is that the disk is established following 
the infall of relatively small amounts of low angular momentum 
gas \citep{king07}. In this scenario, one expects the disk to 
be initially non-circular, with the fluid streamlines describing 
nested aligned ellipses. Partially motivated by such a picture, 
we consider as initial conditions a disk in which $e \neq 0$ and 
$\varpi = {\rm const}$. After time $t$, the pericenters of ellipses 
separated by radial distance $\Delta r$ have angular separation,
\begin{equation} 
 \Delta \varpi = \frac{{\rm d}\omega}{{\rm d}r} \Delta r t.
\end{equation} 
Defining the winding scale as the radial separation for which 
$\Delta \varpi = 2 \pi$,
\begin{equation} 
 \Delta r_{\rm wind} = \frac{4 \pi Q}{c_s} \frac{r^2}{t}. 
\end{equation} 
This scale decreases inversely with $t$, and as a result the 
timescale on which viscosity can damp the eccentricity (equation 
\ref{eq_tcirc}) scales with the system age as,
\begin{equation}
  \tau_{\rm damp} = \frac{\Omega}{\alpha_e c_s^2} 
  \left( \frac{4 \pi Q}{c_s} \right)^2 \frac{r^4}{t^2}.
\end{equation}
The disk will circularize when $t \sim \tau_{\rm damp}$. Defining 
this time as the circularization time $\tau_{\rm circ}$, we obtain,
\begin{equation}
 \tau_{\rm circ} = (4 \pi Q)^{2/3} (GM_{\rm BH})^{1/6} 
 \alpha_e^{-1/3} c_s^{-4/3} r^{5/6},
\end{equation}
which displays only a weak dependence on the black hole mass and 
on the (very uncertain) viscosity damping the eccentricity. Since 
masing action is only possible over a fairly limited range of disk 
temperatures the only significant dependence is the almost linear 
scaling of the circularization timescale with radius. Adopting 
representative values of the parameters the timescale is,
\begin{eqnarray}
  \tau_{\rm circ} &=& 8.5 \times 10^6 Q^{2/3} 
  \left( \frac{M_{\rm BH}}{10^8 \ M_\odot} \right)^{1/6} 
  \left( \frac{\alpha_e}{0.1} \right)^{-1/3} \nonumber \\
  &\times& \left( \frac{c_s}{1.5 \ {\rm kms}^{-1}} \right)^{-4/3}
  \left( \frac{r}{0.1 \ {\rm pc}} \right)^{5/6} \ {\rm yr}.
\label{eq_estimate}  
\end{eqnarray}
This simple calculation provides an estimate of when enough 
differential precession will have accumulated to start eccentricity 
damping. If {\em only} differential precession and damping are 
considered then subsequent evolution will simply wrap up the eccentricity 
vector into an ever-tighter spiral, hastening the decay of ${\cal{E}}^2$ 
even further. Indeed, very rapid damping due to this process was seen 
in numerical solutions for precessing {\em warped} disks in the regime 
where the warp is transmitted viscously \citep{armitage99}. In the eccentric  
case, however, inspection of equation (\ref{eq_simplified}), 
shows that once damping is important (or even beforehand), the wavelike term 
proportional to $\gamma$ is also likely to be significant. This term 
can offset differential precession, and affect the rate at which further 
decay of the eccentricity occurs. As we show subsequently, it turns out 
that while equation (\ref{eq_estimate}) suffices to predict the onset of eccentricity 
damping fairly accurately, the rate of subsequent decay varies with $\alpha_e$ 
much more strongly than would be suspected based on the rather weak scaling 
derived above.

\subsection{Numerical results}
To study the eccentricity evolution in more detail we have computed numerical 
solutions to equation (\ref{eq_simplified}). We discretize the equation on a 
uniform spatial grid between $r_{\rm in}$ and $r_{\rm out}$, and advance the 
system in time using a second-order scheme that, in the absence of damping, preserves the invariant 
${\cal{E}}^2$ to sufficient accuracy for our 
purposes. We assume that the inner disk is circular, and set $E=0$ at the 
inner boundary $r_{\rm in}$. The correct boundary conditions to apply 
at $r_{\rm out}$ are less clear, since the outer edge of the masing 
region may reflect either a physical disk edge (defined, for example, 
by the onset of disk fragmentation) or merely the outer extent of the zone 
that produces maser emission \citep{neufeld94}. Typically we use zero-gradient conditions 
on the components of $E$ (i.e. $\partial E_x / \partial r = \partial E_y / \partial r= 0$ 
at $r_{\rm out}$). We have run models with different boundary conditions, and with different 
choices for $r_{\rm out}$, to gauge the effect of the boundaries on the results.

\begin{figure}
\plotone{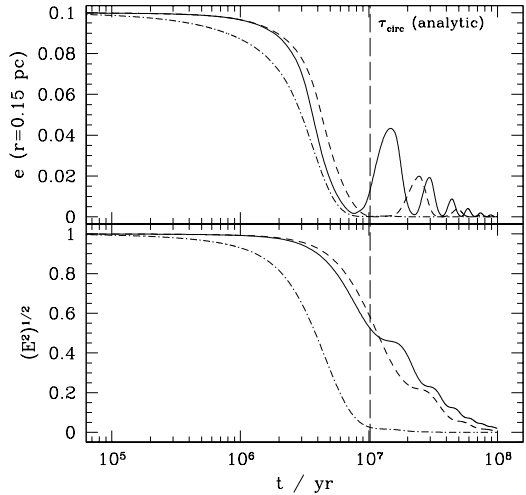}
\caption{The evolution of the eccentricity at $r = 0.15 \ {\rm pc}$ (upper panel) and 
eccentricity invariant $({\cal{E}}^2)^{1/2}$ (lower panel) computed from numerical solutions to 
equation (\ref{eq_simplified}). The units on the vertical axes are arbitrary. Three runs 
are shown, all of which assume a $Q=1$ disk around a a black hole of mass $4 \times 10^7 \ 
M_\odot$. The disk has a constant sound speed of 1.5~km/s and $\gamma = 1.4$. The initial 
eccentricity is a gaussian centered on 0.15~pc with a width of 0.05~pc and a constant $\varpi$. The solid 
curves show results for a disk extending from 0.05~pc to 0.4~pc, with $\alpha_e = 0.1$. 
The dashed curves depict the evolution of an identical disk within larger boundaries that 
extend from 0.025~pc to 0.8~pc. The vertical dashed line shows the analytic estimate for 
the circularization time for these parameters. The dot-dashed curves show the effect of 
increasing the damping coefficient to $\alpha_e = 1$.}
\vskip0.2truein
\label{f1}
\end{figure}

Figure~\ref{f1} shows the decay of an initially gaussian eccentricity perturbation,
\begin{eqnarray}
 e(r) & = & e_0 \exp \left[ - \frac{(r-r_0)^2}{\Delta r^2} \right] \nonumber \\
 \varpi(r) & = & 0
\end{eqnarray}
centered on $r_0 = 0.15$~pc with width $\Delta r = 0.05$~pc. As this is a linear 
calculation, the value of $e_0$ is entirely arbitrary. We assume a $Q=1$ disk 
with a constant sound speed $c_s = 1.5$~km/s, which results in a 
$\Sigma \propto r^{-3/2}$ surface density profile. The black hole mass is 
set to the NGC~4258 value of $4 \times 10^7 \ M_\odot$. The Figure shows 
results obtained from two models with $\alpha_e = 0.1$ (which differ only 
in the radial extent of the disk), and one model with stronger eccentricity 
damping $\alpha_e = 1$. We plot both the eccentricity of the disk at $r_0$, 
and the integrated eccentricity invariant $\sqrt{{\cal{E}}^2}$ in a form that is 
proportional to the magnitude of the eccentricity.

The numerical results confirm that the analytic estimate of the damping timescale 
(equation \ref{eq_estimate}) is reasonably accurate. For the two runs with 
$\alpha_e = 0.1$ we find that $\sqrt{{\cal{E}}^2}$ drops by a factor of two 
after 10~Myr of evolution, as predicted. For these runs, however, subsequent 
damping is relatively slow. It takes approximately 50~Myr for $\sqrt{{\cal{E}}^2}$ 
to decay from its initial value by an order of magnitude. The corresponding curve showing the 
local value of $e$ displays clearly the wave-like nature of the evolution. Although $e$ 
displays a general decaying trend the evolution is oscillatory rather than a 
monotonic decline, with the period of the oscillations varying with the radial 
extent of the disk. Only for strong damping ($\alpha_e \approx 1$) is the 
wave-like behavior suppressed. In the strong damping limit the initial decay 
of the eccentricity is not that much faster than in the weakly damped cases 
but, once damping has set in, the subsequent decline of the eccentricity to 
negligible values is rapid.

We have also studied the effect of different initial conditions. In general, we 
find that the decay rate of $\sqrt{{\cal{E}}^2}$ is quite robust against 
changes in the initial conditions. The evolution of the value of $e$ at a 
particular radius, on the other hand, can exhibit large transients at relatively 
early times if the initial conditions contain large gradients in $E$. Physically 
it seems likely that an eccentric disk would settle down to a smooth distribution 
of $E$ shortly after formation, so the gaussian initial conditions we have plotted 
ought to be representative of the likely evolution.

\begin{figure*}
\plotone{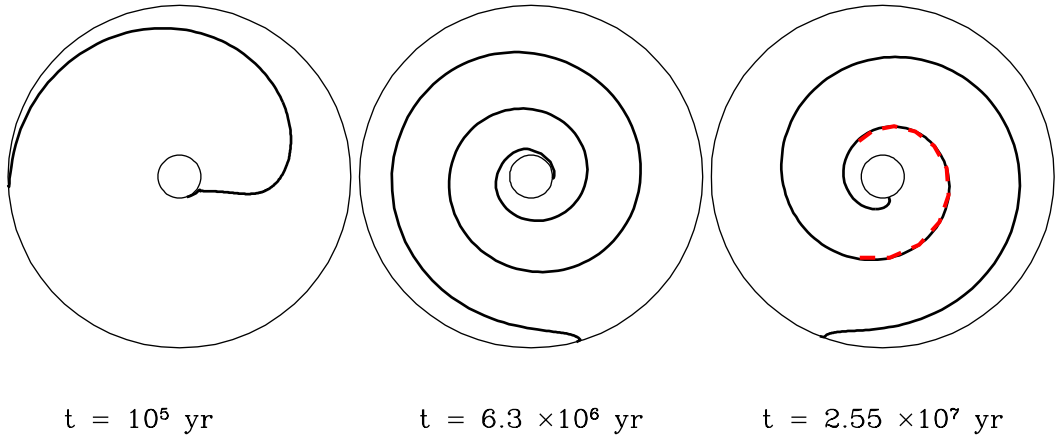}
\vskip-3.2truein
\caption{The radial variation of the angle of pericenter $\varpi$ of the fluid streamlines, 
plotted at different times. The run corresponds to the solid curves in Fig.~\ref{f1}. The 
eccentricity vector is wound up by differential precession until damping sets in. As waves 
subsequently propagate through the disk the spiral pattern periodically reverses, as shown 
in the rightmost panel. The heavy dashed curve in the rightmost panel shows a linear fit 
to $\varpi(r)$, which is reasonably accurate at most epochs over a limited range (a factor 
of two) in radius.}
\label{f2}
\vskip0.2truein
\end{figure*}

\begin{figure}
\plotone{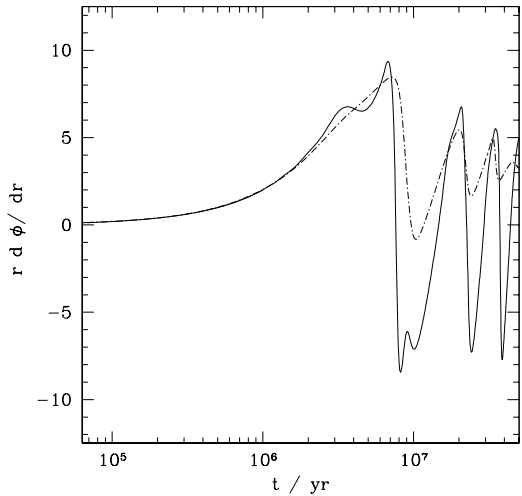}
\caption{The evolution of the twist $r {\rm d}\varpi / {\rm d r}$ in the eccentricity vector, 
evaluated at $r = 0.15$~pc for the standard run (solid curve, with parameters corresponding 
to the same curve in Fig.~\ref{f1}) and the run with $\alpha_e = 1$. The weakly damped case 
displays approximately periodic reversals between leading and trailing spiral patterns in the 
eccentricity vector. These occur on a timescale of the order of $10^7$~yr. The oscillations 
are largely suppressed in the run with stronger eccentricity damping.}
\vskip0.2truein
\label{f3}
\end{figure}

Figure~\ref{f2} plots the spiral traced by $\varpi(r)$ at different times in the 
weakly damped run ($\alpha_e = 0.1$) with the smaller radial extent of the disk 
($0.05 \ {\rm pc} < r < 0.4 \ {\rm pc}$). The initial evolution of the eccentricity 
vector is driven by differential precession, with damping setting in once the spiral 
has become sufficiently tightly wound -- for this run a value of $r {\rm d}\varpi / {\rm d}r 
\simeq 2 \pi$, which is achieved after a few Myr, suffices. During the damping phase 
there is no further increase in the winding due to differential precession, which is 
offset by the wave-like term in the evolution equation. As a consequence, final 
damping of the eccentricity in the low $\alpha_e$ runs occurs relatively slowly. 
Indeed, as shown in Figure~\ref{f3}, for the low $\alpha_e$ runs the sense of the 
spiral pattern {\em reverses} during the damping phase roughly periodically, though 
only the first of these reversals is likely to have any physical significance since at 
later times the actual magnitude of the eccentricity is quite small. As with the 
eccentricity oscillations described above, the reversals are largely eliminated 
in the more strongly damped run.

Based on the above results, it is clear that while the eccentricity is decaying 
the structure of the disk can be complex. It is not even possible to predict 
securely whether the eccentricity vector will describe a leading or a trailing 
spiral pattern. At almost all epochs a description of the disk in 
terms of aligned nested ellipses is a very poor approximation. A better 
description is to assume a linear radial variation in $\varpi$,
\begin{equation}
 \varpi(r) = \varpi(r_0) + \left( \frac{{\rm d}\varpi}{{\rm d}{r}} \right)_{r_0} (r-r_0).
\end{equation} 
As shown in Figure~2, this description is usually reliable over at least a factor of 
two in radius. Typical values of the twist in the eccentricity vector, 
$r {{\rm d}\varpi} / {{\rm d}{r}}$ are in the range of $\pi$ to $2 \pi$. We believe that 
fitting a model of this kind to observational data would represent a substantial 
improvement over a model with $\varpi = {\rm const}$, though at the cost of 
introducing an additional free parameter in the winding angle of the eccentricity 
vector.

\section{Discussion}
Theory provides scant guidance as to the formation mechanism and lifetime 
of sub-pc disks in AGN. The long viscous timescale at these radii -- of 
the order of a Gyr -- means that it is unlikely that such disks formed 
from outward viscous expansion of gas that was initially much closer to the black hole, but 
this still leaves several possibilities for feeding gas directly to sub-pc 
radii. One idea 
is that clouds on low angular momentum orbits are sporadically shredded 
by the tidal field of the black hole, with the debris cooling and 
settling into a disk \citep{king07}. Depending upon the disk mass and 
radius involved, the result could be a stable gaseous disk or gravitational 
fragmentation into a disk of stars \citep{gammie01,rice03}. Such an idea provides one 
possible explanation for the disk of young stars observed at sub-pc radii 
from our own Galactic Center \citep{levin03}. If this is the origin 
of the masing disks then it is probable that the initial conditions would involve eccentric 
and / or warped disks. However, there are also plausible scenarios 
that would yield initially circular disks. For M31, for example, it 
has been suggested that the innermost, circular, disk of stars formed 
from a gravitationally unstable disk that accumulated from the 
stellar winds of stars situated further out \citep{chang07}. Given 
these very different models, the observation that the masing disk in NGC~4258 
is at least approximately circular \citep{humphreys08} raises two 
questions. First, does the lack of obvious eccentricity constrain 
the formation mechanism to favor the scenarios that yield initially 
circular orbits? Second, it is reasonable to assume that the 
(unmeasured) eccentricity is actually vanishingly small? In this 
paper we have attempted to provide partial answers to both of these questions.

Our main result is that the {\em initial} damping of a thin self-gravitating 
disk at sub-pc distances from a black hole is relatively rapid. The timescale 
at 0.1~pc is around 10~Myr, and even this may be an overestimate since any 
nonlinear effects (such as the formation of shocks as elliptical orbits precess) 
will hasten the decay further. Rapid damping is driven by differential precession, 
which occurs on a much shorter timescale than the viscous evolution of the disk. 
The lack of gross eccentricity in NGC~4258 is therefore not surprising, and 
does not constrain the disk formation mechanism. Once damping is underway, 
however, it can take a surprisingly long period for the eccentricity to 
decay to negligible levels. If the viscous coefficient governing eccentricity 
damping is relatively small ($\alpha_e = 0.1$ in our description) the presence 
of waves in the disk allows significant eccentricity to survive for up to 50~Myr. 
Evidently it would be useful to {\em determine} the efficiency of eccentricity 
damping under the conditions likely to prevail at sub-pc  
radii, perhaps via improved simulations of self-gravitating eccentric disks, which 
to date have focused on simpler questions such as whether the disk fragments or 
not \citep{alexander08}. Currently, though, it seems unwise to conclude that a 
good upper limit on the eccentricity implies strictly circular orbits.

Finally, we addressed the question of the predicted orbital structure of an 
eccentric disk during the circularization process. Existing constraints on the 
eccentricity of the NGC~4258 disk \citep{humphreys08} have been derived assuming 
a model, developed by \cite{statler01} for different purposes, in which the 
eccentric orbits are aligned and nested. In the case of very thin masing 
disks this model is not a good approximation to the likely structure, which 
instead involves a moderately tightly wound spiral in the angle of periapse. 
A reasonable fit, over a modest range of radii, is possible using the next 
order approximation in which the periapse angle is a linear function of 
orbital radius.

\acknowledgements

This work was stimulated by discussions with Fred Lo, and completed at the 
urging of Dick McCray, to whom I am indebted. My research was supported by NASA 
under grants NNG04GL01G and NNX07AH08G from the Astrophysics Theory Programs, and 
by the NSF under grant AST~0407040.


\begin{thebibliography}{}

\bibitem[Alexander et al.(2008)]{alexander08}
 Alexander, R. D., Armitage, P. J., Cuadra, J., \& Begelman, M. C. 2008, \apj, in press (arXiv:0711.0759v1) 

\bibitem[Argon et al.(2007)]{argon07}
 Argon, A. L., Greenhill, L. J., Reid, M. J., Moran, J. M., \& Humphreys, E. M. L. 2007, 
 \apj, 659, 1040

\bibitem[Armitage \& Natarajan(1999)]{armitage99}
 Armitage, P. J., \& Natarajan, P. 1999, \apj, 525, 909

\bibitem[Balbus \& Hawley(1998)]{balbus98}
 Balbus, S. A., \& Hawley, J. F. 1998, Reviews of Modern Physics, 70, 1

\bibitem[Chang et al.(2007)]{chang07}
 Chang, P., Murray-Clay, R., Chiang, E., \& Quataert, E. 2007, \apj, 668, 236

\bibitem[Clarke(1988)]{clarke88}
 Clarke, C. J. 1988, \mnras, 235, 881

\bibitem[Elitzur, Hollenbach \& McKee(1989)]{elitzur89}
 Elitzur, M., Hollenbach, D. J., \& McKee, C. F. 1989, \apj, 346, 983

\bibitem[Gammie(2001)]{gammie01}
 Gammie, C. F. 2001, \apj, 553, 174

\bibitem[Goodchild \& Ogilvie(2006)]{goodchild06}
 Goodchild, S., \& Ogilvie, G. 2006, MNRAS, 368, 1123

\bibitem[Goodman(2003)]{goodman03}
 Goodman, J. 2003, \mnras, 339, 937

\bibitem[Herrnstein et al.(1999)]{herrnstein99}
 Herrnstein, J. R., Moran, J. M., Greenhill, L. J., Diamond, P. J., Inoue, M., Nakai, N., Miyoshi, M., Henkel, C., 
 \& Riess, A. 1999, Nature, 400, 539
 
\bibitem[Hu(2005)]{hu05}
 Hu, W. 2005, in Observing Dark Energy, eds S.~C. Wolff \& T.~R. Lauer, ASP Conference Series, Vol. 339, 
 ASP (San Francisco), p.~215 (arXiv:astro-ph/0407158v1)
 
\bibitem[Humphreys et al.(2008)]{humphreys08}
 Humphreys, E. M. L., Reid, M. J., Greenhill, L. J., Moran, J. M., \& Argon, A. L. 2008, 
 ApJ, 672, 800

\bibitem[Kato(1983)]{kato83}
 Kato, S. 1983, PASJ, 35, 249

\bibitem[King \& Pringle(2007)]{king07}
 King, A. R., \& Pringle, J. E. 2007, \mnras, 377, L25

\bibitem[Levin \& Beloborodov(2003)]{levin03}
 Levin, Y., \& Beloborodov, A. M. 2003, \apj, 590, L33

\bibitem[Kolychalov \& Sunyaev(1980)]{kolychalov80}
 Kolychalov, P. I., \& Sunyaev, R. A. 1980, Soviet Astronomy Letters, 6, 357 

\bibitem[Maoz(1995)]{maoz95}
 Maoz, E. 1995, \apj, 455, L131
 
\bibitem[Maoz \& McKee(1998)]{maoz98}
 Maoz, E., \& McKee, C. F. 1998, \apj, 494, 218 

\bibitem[Miyoshi et al.(1995)]{miyoshi95}
 Miyoshi, M., Moran, J., Herrnstein, J., Greenhill, L., Nakai, N., Diamond, P., \& Inoue, M. 1999, 
 Nature, 373, 127
 
\bibitem[Neufeld, Maloney \& Conger(1994)]{neufeld94}
 Neufeld, D. A., Maloney, P. R., \& Conger, S. 1994, \apj, 436, L127
 
\bibitem[Ogilvie(2001)]{ogilvie01}
 Ogilvie, G. I. 2001, \mnras, 325, 231 
 
\bibitem[Olling(2007)]{olling07}
 Olling, R. P. 2007, MNRAS, 378, 1385 
  
\bibitem[Rice et al.(2003)]{rice03}
 Rice, W. K. M., Armitage, P. J., Bate, M. R., \& Bonnell, I. A. 2003, \mnras, 339, 1025 
  
\bibitem[Shakura \& Sunyaev(1973)]{shakura73}
 Shakura, N. I., \& Sunyaev, R. A. 1973, 
 A\&A, 24, 337
 
\bibitem[Shlosman, Begelman \& Frank(1990)]{shlosman90}
 Shlosman, I., Begelman, M. C., \& Frank, J. 1990, Nature, 345, 679
 
\bibitem[Statler(2001)]{statler01}
 Statler, T. S. 2001, \aj, 122, 2257
 
\bibitem[Toomre(1964)]{toomre64}
 Toomre, A. 1964, \apj, 139, 1217 
  
\end{thebibliography}
\end{document}